\documentstyle[12pt]{article}
\newcommand{\beq}{\begin{equation}}
\newcommand{\eeq}{\end{equation}}
\newcommand{\beqa}{\begin{eqnarray}}
\newcommand{\eeqa}{\end{eqnarray}}
\newcommand{\ba}{\begin{array}}
\newcommand{\ea}{\end{array}}

\begin{document}

\begin{flushright}
Preprint CAMTP/99-1\\
March 1999\\
\end{flushright}

\vskip 0.5 truecm
\begin{center}
\Large
{\bf  Study of regular and irregular states in generic systems}\\
\vspace{0.25in}
\normalsize
Gregor Veble, Marko Robnik and Junxian Liu \footnote{e--mails: 
robnik@uni-mb.si, gregor.veble@uni-mb.si. junxian.liu@uni-mb.si}\\
\vspace{0.3in}
Center for Applied Mathematics and Theoretical Physics,\\
University of Maribor, Krekova 2, SI--2000 Maribor, Slovenia\\

\end{center}

\vspace{0.3in}

\normalsize
\noindent
{\bf Abstract.} In this work we present the results of a numerical and 
semiclassical analysis of high lying states in a Hamiltonian system,
 whose classical mechanics is of a generic, mixed type, where the energy 
surface is split into regions of regular and chaotic motion. As predicted by
 the principle of uniform semiclassical condensation (PUSC), when the effective
$\hbar$ tends to $0$, each state can be classified as regular or irregular.
We were able to semiclassically reproduce individual 
regular states by the EBK torus quantization, for which we devise a new 
approach, while for the irregular ones we
found the semiclassical prediction of their autocorrelation function,
in a good agreement with numerics. We also 
looked at the low lying states to better understand the onset 
of semiclassical behaviour.

\vspace{0.6in}

\\\\
Submitted to {\bf Journal of Physics A: Mathematical and General}
\normalsize
\vspace{0.1in}
  
\newpage

\section{Introduction}

\noindent                     
There is a remarkable difference between the quantum properties of the 
classically integrable and fully chaotic (ergodic) systems. 
While in the integrable case
the wavefunctions possess an ordered structure, the eigenstates of classically
ergodic systems appear random (Berry 1977, Voros 1979) 
in the semiclassical limit $\hbar
\to 0$. They can be locally well represented as a superposition of plane waves
with equal wavevector magnitude but random phases, leading to the Gaussian
distribution of the wavefunction amplitude. The assumption of random phases
may, however, 
break down on dynamical grounds when $\hbar$ is not sufficiently small.
This leads to the
phenomenon of scars, which are the regions of amplified wavefunction amplitude
close to the short and weakly unstable classical periodic orbits (Heller 1984).

In the classically mixed systems the energy (hyper)surface is split into
both chaotic regions, within which the motion is ergodic, 
and the regular regions where motion is, as in the fully integrable case,
quasiperiodic and confined to invariant tori within the energy surface. 
We call such
systems generic since this is the most general type of Hamiltonian dynamics.
The principle of uniform semiclassical condensation (PUSC, see Robnik 1988,1998)
states that the quantal phase space distribution of any eigenstate, 
given by its Wigner function, should be uniformly distributed on a
classically invariant object in the phase space when $\hbar \to 0$. 
This object can be either an
invariant torus in one of the regular regions in the phase space 
or a whole chaotic component.
   
While the PUSC is valid for fully integrable and ergodic systems, its full
potential is shown when applied to the generic case. Here it predicts that the
eigenstates are separated into regular and irregular ones, depending on whether
the classical object onto which their Wigner function condenses is a regular
torus or a chaotic component, respectively. This has far reaching consequences
leading, for example, 
to the picture of Berry and Robnik (1984) for the statistics of
energy levels, where the regular and chaotic states are assumed to contribute
independent level sequences to the total spectrum. 
This was confirmed by many numerical
computations (Prosen and Robnik 1993b, 
1994 and 1999, Prosen 1995, 1996 and 1998, Robnik 1998).

In this work we are interested in geometrical and statistical properties 
of high lying eigenfunctions. It is an extension of previous work done on mixed
type systems by Prosen and
Robnik (1993c,1994) and by Li and Robnik (1995a,b), 
but also relates to and further develops the
paper by Li and Robnik (1994) concerning the statistical properties of chaotic
states, which drastically differ from the properties of regular
states in classically integrable systems (Robnik and Veble 1998).
The main step forward in the numerical direction was the use of the
so-called scaling method, introduced by Vergini and Saraceno (1995), with which
we obtained the states with consecutive indices around $2.5\cdot 10^6$
and thus it enabled us to go much farther into the semiclassical region. On
the other hand, we were able to semiclassically reconstruct the
numerically obtained regular states
by employing the Einstein-Brillouin-Keller (EBK) torus quantization 
in section 3.1, where we offer a new approach to this problem, especially
on the numerical side.
While we were not able to semiclassically 
reconstruct the chaotic
states, we obtained a good prediction of their statistical properties by
strictly employing the PUSC (see Robnik 1998) in section 3.2. 

\section{Our catalogues of states}

We dealt with a model billiard system obtained by conformally mapping the unit
circle with the complex quadratic polynomial, as introduced by Robnik 
(1983, 1984),

\beq
 z \to w(z) = z + \lambda z^2,\ w(z)=x + i y.
\eeq 

The range of $x$ at $y=0$ inside the billiard is $x \in 
[-1+\lambda,+1+\lambda]$.
We used the value of $\lambda=0.15$, where the classical phase space is roughly
equally divided into components of regular and chaotic motion. We chose the
Poincar\' e surface of section (SOS) to lie 
on the symmetry axis $y=0$ with coordinate $x$
and the conjugate momentum $p_x$ as the parameters of the surface. The
intersection of the main
chaotic component of our billiard with the SOS is shown in the figure 
\ref{sos}. The coordinate $x$ is taken 
{\em relative to the center of the billiard} (i.e. it is shifted
by $\lambda$ w.r.t. $x=0$ of equation (1), so that now
the range of $x$ is $x\in [-1,+1]$), while 
$p_x$ is the $x$-component of the unit momentum vector.

The quantum
mechanics of billiards is described by the Helmholtz equation
\beq
(\Delta + k^2) \psi=0,
\eeq
with the Dirichlet boundary conditions, where $k^2=2mE/\hbar^2$.
We limited ourselves to the states with even parity with respect to reflection
across the symmetry line $y=0$. 

For each state we calculated the smoothed 
projection of the Wigner function. The Wigner function of a state 
$\psi({\bf q})$ in the general case of $N$ degrees of freedom
is defined in the full phase space $({\bf q},{\bf p})$ as (Berry 1983)
\beq
  W({\bf q},{\bf p})=\frac{1}{(2\pi\hbar)^N}\int d^N 
 {\bf X} \exp(-i {\bf p}\cdot
  {\bf X}/\hbar)\psi^{\dagger}({\bf q}-{\bf X}/2)\psi({\bf q}+{\bf X}/2).
  \label{Wigner}
\eeq
In our case the eigenfunctions $\psi (x,y)$ generate their Wigner transforms 
$W(x,y,p_x,p_y)$ through (\ref{Wigner}), where $N=2$.
In order to compare the Wigner function of a state of our system 
with the classical SOS plot we took its value on the 
symmetry line ($y=0$) and integrated it over $p_y$,
\beq
  \rho_{SOS}(x,p_x)=\int dp_y W(x,y=0,p_x,p_y).
\eeq
The result is
\beq
  \rho_{SOS}(x,p_x)=\frac{1}{2\pi\hbar}\int dX \exp(-i p_x X/\hbar) 
  \psi^{\dagger}(x-X/2, y=0) \psi(x+X/2, y=0).
\eeq
Here we see the reason for considering the even parity states only, because 
$\psi(x,y=0)$ is exactly zero for odd states, and therefore a different approach
must be used to analyze them.

As is well known, the Wigner function is not positive definite but exhibits 
small oscillations that can blur the overall picture. We chose to smooth the
projection of the Wigner function by a suitable Gaussian. It was chosen 
narrower
than the minimum uncertainty Gaussian (in which case the Wigner function 
becomes the positive definite Husimi distribution) in order not to smooth out
too many features, but still wide enough to reduce the oscillations.

The first catalogue of eigenstates and the corresponding smoothed Wigner 
function projections comprises the first 1000 even states. 
They were obtained by
the conformal mapping diagonalization technique (as described in Robnik 1984).
We stress
that by this method no levels and eigenstates were lost, 
as can happen with other approaches
such as Heller's plane wave decomposition method (Heller 1984)
and/or the boundary integral method (see e.g. Berry and Wilkinson 1984) etc.
 
Such a complete catalogue 
gives us a good picture of the overall behaviour of the system. 
Many of the states in this low energy region
can be associated with the shortest classical periodic orbits. In a mixed type
system such as ours, there are both stable orbits that 
are found within the islands
of stability, or unstable ones that lie within one of the chaotic components.

The shortest periodic orbits of our system are shown in figure \ref{orbits}.
These are the stable (labeled by 1) and unstable (5) periodic orbit with two
bounces, the stable (2) and unstable (6) three bounce periodic orbit and the
stable periodic orbit with four bounces (4).
As our billiard is convex, there also exists an infinity of
 stable periodic orbits skipping
along the boundary (3) of the billiard, 
which are associated with and support the whispering gallery modes 
(Lazutkin 1981,1991, Li and Robnik 1995a).

Examples of the states that correspond to the stable 
periodic orbits with the indices (1-4) in the figure \ref{orbits} are shown in
the corresponding rows of the 
figure \ref{lowstatesstable} with the smoothed projections of their
Wigner functions shown in the figure \ref{lowstateswignerstable}. Each row 
shows the states of the same type with increasing energy. 
The states corresponding to the stable periodic orbits appear consistently
and systematically
across the catalogue of states. As we will see later, they can be attributed to
the quantized tori in the regular regions, and can be found for those tori
whose actions satisfy the Einstein-Brillouin-Keller 
quantization condition (see Robnik 1998). If we compare the Wigner 
plots of these states with the SOS plot in figure \ref{sos}, we notice that 
the areas of 
greatest intensities of Wigner functions 
are found within the corresponding islands of stability in the SOS plot.

The states that correspond to the unstable periodic orbits with indices (5) 
and (6)
are shown in the rows of the 
figure \ref{lowstatesunstable} with the corresponding smoothed projections
of their Wigner functions given in the figure \ref{lowstateswignerunstable}, 
with their energy again increasing along the rows.
The states
corresponding to the unstable periodic orbits emerge with varying
intensities with respect to the background, and are less frequent with
increasing energy. The approximate 
position of their emergence in the spectrum can be
determined by the condition that the classical action along the periodic orbit
should be a multiple of Planck's constant plus the Maslov's phase corrections
due to the caustics formed by nearby trajectories (Robnik 1989). 
Therefore, in 2-dim billiards, 
the intervals (either in energy $E$ or the cumulative number of 
states ${\cal N}$) of consecutive re-appearance of the eigenstates 
of the same type grow as the square root of $E$ or ${\cal N}$. 
Such states with increased intensities of amplitude close to the classically
unstable periodic orbits are called scars (Heller 1984, 1986, Bogomolny 1988).

With increasing energy the 
longer periodic orbits start to manifest themselves in the structure
of eigenstates, however, many of the states are becoming 
increasingly difficult 
 to associate with simple periodic orbits. Two examples of such
states are shown in figure \ref{twostates}.
The left state is a regular state spanned by a torus in the neighbourhood of
the stable periodic orbit with three bounces, characterised by two quantum
numbers. It is interesting to note that
while the quantum number along the direction of the periodic orbit is quite
large (about $140$), the transversal quantum number is equal to $1$,
giving rise to a single nodal line along the direction of the periodic orbit.
On the other hand the state on the right can not be associated  only with a
single unstable periodic orbit, but with a large portion of the chaotic
component, this being more prominent in the smoothed projections of the 
corresponding Wigner
function shown in the lower row of the figure \ref{twostates}.

The separation of states into regular and irregular ones becomes 
fully explicit in our second catalogue of states.  
The catalogue consists of $100$
consecutive states starting at the consecutive index of about $2.5\cdot 10^6$.
These states were obtained by the scaling method first introduced by Vergini
and Saraceno (1995), that enables us to find a few states in the neighbourhood
of a chosen wavenumber $k$. As this is a diagonalizational method no
levels were missed. Almost each level in our small catalogue can be clearly
identified as regular or irregular, an idea proposed already by Percival
(1973). The only exception in the catalogue is a pair of states lying close 
together with respect to the mean level spacing, 
shown in figure
\ref{pair}, where both of the states are superpositions of a regular ($|\psi_r
\rangle$) and irregular ($|\psi_i\rangle$) state. These two states
are close to a degeneracy of two energetically equal but structurally
different quantized classical objects.

This mixing is the consequence of the fact that the regular and irregular
state are not exact solutions of the Hamiltonian which leads to the matrix 
element
\beq
H_{ri}=\langle \psi_r|\hat H |\psi_i\rangle
\label{matel}
\eeq
not being equal to $0$. When two such states, or, more precisely, their 
adiabatically corresponding states, are brought close together on the energy 
scale by varying a parameter of the system (e.g. $\lambda$ in our billiard),
their eigenenergies do not cross but show a phenomenon of level repulsion
(avoided crossing). In
the cases when the effect of other levels can be neglected, the smallest 
energy spacing between the two levels reached is twice the value of the matrix 
element (\ref{matel}). At this point the eigenstates are exactly the symmetric
and antisymmetric superpositions of the well-separated states. If we vary the 
parameter of the system further, the adiabatic equivalents of the original
states will exchange identities. 

While the avoided level crossings are typical for irregular states, the 
regular states do not exhibit the phenomenon of level repulsion amongst 
themselves (Berry 1983), except possibly on an exponentially
small scale (due to the tunneling effects). 
Furthermore, according to the PUSC, there should be no level repulsion between
the regular and irregular states as $\hbar \to 0$. As this repulsion is 
directly connected to the mixing of states, the relative number of mixed 
states, such as the pair shown in figure \ref{pair}, is expected to
tend to $0$ when effective Planck's constant of the system tends to $0$.

More states of both the regular and irregular type from this catalogue
 with their appropriate analysis will be shown in the next section.

\section{Analysis of states}

The main purpose of our work was to understand to what extent it is possible
to describe the structure of individual eigenstates by semiclassical methods.
For completely integrable systems there exists the Einstein-Brillouin-Keller
method of torus quantization (see for example Berry 1983, Robnik 1998), 
where each semiclassical 
eigenstate is spanned by an invariant torus in classical phase space, for which
the classical actions are integer multiples of $\hbar$ with corrections due to
the singularities of projection of the torus onto the configuration space.
For fully chaotic systems, and general mixed systems (generic systems), 
the Gutzwiller periodic orbit theory  (Gutzwiller 1990) can in principle
be employed
to obtain the density of states (the spectrum) and the wavefunctions,
to the leading semiclassical approximation.

The semiclassical methods cannot, however, predict individual energy levels
within a vanishing fraction of the mean level spacing even in the limit 
$\hbar\to 0$ (Prosen and Robnik 1993a, Robnik and Salasnich 1997). 
This limits their use in the analysis of statistical 
properties of spectra. Furthermore, in the chaotic case the levels are also 
very sensitive to perturbations of the system (Percival 1973). 
The Gutzwiller approach and method is very useful in the qualitative 
analysis, and in certain context also quantitative (in describing
the collective and statistical properties), but it is still just the 
leading term in a certain semiclassical
expansion, not good enough to resolve the fine structures with sufficient
accuracy to make the analysis of {\em individual states and energy levels}
reliable. Therefore, because of the sensitivity of chaotic eigenstates
and the approximating nature of the theory, to some extent 
the questions about the fine structure of individual chaotic 
states are irrelevant. It is therefore 
 more appropriate to discuss the statistical properties of chaotic 
states, which are, however, less sensitive to perturbations.

In a mixed system the phase space is divided into chaotic and regular 
components. Our work was guided by the principle of uniform semiclassical 
condensation (PUSC, see Robnik 1998), stating that when $\hbar$ tends to $0$ 
the Wigner function of any eigenstate uniformly condenses on an invariant 
object in phase space. This can be either a torus in the regular region or a
whole chaotic component. Each state could thus be
labeled as either regular or irregular (chaotic) in the semiclassical limit.
By looking at the catalogue of states at high (and to some extent even at low)
energies, one can see that this can indeed be done, though there is
still the localization phenomenon present due to the still insufficiently low 
value of the effective Planck's constant.

\subsection{Regular states}
\label{analysis}

We start the analysis by considering 
the regular states. These are the states that can be
attributed to quantized tori within the regular regions. 
For these states we tried to employ the EBK torus quantization. We construct
a wavefunction on the torus as a sum of contributions
\beq
  \psi_j({\bf q})=A_j({\bf q})\exp(\frac{i}{\hbar}S_j^{cl}({\bf q})+\phi_j)
  \label{semiclassical}
\eeq                   
of different projections $j$ of the torus onto configuration space.
$S_j$ is the classical action with respect to some point on the torus and
$A_j^2$ the classical density of trajectories on this projection.
The phase
of the wavefunction must change by an integer multiple of $2 \pi$
when going around any closed contour of the torus. 
This gives us the quantization conditions
\beq
  I_i= \frac{1}{2\pi} 
   \oint_{\gamma_i}{\bf p}\cdot d{\bf q}=\hbar (n_i+\beta_i/4)
  \label{quantization}
\eeq
where $\gamma_i$ are the irreducible closed contours on the torus and $n_i$ the
torus quantum numbers. The integers $\beta_i$ are Maslov's corrections and 
arise
due to the changes of phase $\phi_j$ at the singularities of projection of 
the torus onto 
configuration space. At each caustic encountered along the contour $\gamma_i$
the wavefunction acquires a negative phase shift of 
$\pi/2$, and shifts by $\pi$ when reflected from a hard 
wall\footnote{If the contour passes the singularity in the contrary
direction to that of the Hamiltonian flow on the torus, the phase shifts are of
the opposite sign}. 
From this consideration it follows that $\beta_i$ counts the number of 
caustics plus twice the number of hard walls encountered along the contour.

The main problem arises since, unlike in many completely integrable cases, the 
transformation to action-angle variables for regular components of mixed type 
systems is usually not known. The object we are dealing with is only the 
numerically calculated trajectory, so we must make the best of it.

The task of finding the semiclassical EBK wavefunctions can be divided into 
two 
parts. The first one is finding the torus with the desired quantum numbers 
$n_i$, the second one being the construction of its appropriate wavefunction
in configuration space.

In our billiard system all the regular tori except the whispering
gallery ones wind around a stable periodic orbit with a finite number of
bounces $l_b$. We choose $\theta_1$ to represent the
 movement of the trajectories
along the corresponding periodic orbit while $\theta_2$ represents 
the winding of the trajectories around it. The Maslov index for the contour 
along $\theta_2$ is $\beta_2=2$ since there are two singularities of
projection (caustics) encountered, while for the contour along $\theta_1$
this index is equal to $\beta_1=2 l_b$. In the case of whispering gallery
modes one similarly obtains $\beta_1=0$ and $\beta_2=3$, if $\theta_1$ is 
taken along the boundary of the billiard and $\theta_2$ 'perpendicular' to it.
                                         
We may find the appropriate torus by iteration.  We start with a trajectory 
$\eta$ in a regular island on the SOS and follow it sufficiently long until 
it returns to the desired neighbourhood of 
the initial point, thus approximately completing $N_1$ integer number of 
cycles in $\theta_1$ and $N_2$ in 
$\theta_2$ in the time $T$. 
One can obtain the numbers $N_1$ and $N_2$ by knowing the total 
number of bounces $N_b$ and $N_c$ of the caustics encountered during the 
process through the topological properties of the torus. For a torus 
winding around a periodic orbit, $N_1=N_b/l_b$ and $N_2=N_c/2$. 
The winding frequencies on the torus are then given by
\beq
  \omega_i=2\pi N_i/T.
\eeq               

Finding
the number of bounces $N_b$ along the trajectory is straightforward. 
The number of caustics $N_c$ may be obtained by starting  a trajectory 
$\eta^\prime$ near the original trajectory $\eta$ 
on the same torus (more are needed in case of more than two
dimensions). Whenever the two trajectories cross there is a singularity of the
density of trajectories and hence a caustic. 

One could of course use the
monodromy matrix to find the caustics. There are two reasons for
not doing so. The first one is the numerical simplicity of our approach. The
more important reason is, however, 
that if the trajectories $\eta$ and $\eta^\prime$ 
are taken too close together (infinitesimally
close together in the monodromy matrix approach) we may observe caustics due to
the possible graining of the desired torus into smaller islands of stability,
which quantum mechanics is at the given value of effective Planck's constant
still unable to resolve. A good criterion is that
the trajectories should be separated by $\approx 1/n_i$ for each $\theta_i$

We still have to calculate the action integrals on the torus. The integral
along $\theta_2$ can be calculated by the integral 
\beq
  I_2=\frac{1}{2 \pi} \oint_{\gamma_2} {\bf p} \cdot d{\bf q}
\eeq
along the curve $\gamma_2$ 
formed by the crossing points of the trajectory with the SOS. 
This integral is just $1/2\pi$ of 
the area of the intersection of the torus with the SOS. 
Another action integral we can calculate is the action along the
chosen orbit $\eta$,
\beq
  I=\frac{1}{2\pi}\int_{\eta} {\bf p} \cdot d{\bf q}.
\eeq
This integral is the sum of $I=N_1 I_1+ N_2 I_2$, so the integral along
$\theta_2$ is equal to
\beq
  I_1=(I-N_2 I_2)/N_1.
\eeq
We must iterate this procedure by choosing different starting points 
until the proper torus fulfilling the conditions (\ref{quantization})
has been found. In general there are as many conditions as there are degrees
of freedom. For two dimensional billiard systems whose dynamics is
independent of energy, all action integrals can be written in the form
\beq
 I_{m}=G_m \sqrt{E},
\eeq
where $G_m$ is an integral dependent purely upon 
the geometry of the classical object in question. In order to obtain the 
correct quantized torus, instead of solving two separate equations, due to
this scaling property we 
need only to find the appropriate ratio of the (geometric)
actions 
\beq
  \frac{I_1}{I_2}=\frac{n_1+\beta_1/4}{n_2+\beta_2/4}.
\eeq
We used the robust bisection method to fulfill this condition 
since the dependence of actions upon
initial conditions is not smooth due to the previously mentioned graining of
the tori. Note that this procedure yields only the correct geometry of the 
quantized torus, the energy of which must still be determined by the
quantization conditions \ref{quantization}.

The classical action as a function of time can then be written as
\beq
  \int_{{\bf q}_0}^{{\bf q}(t)} L({\bf q},\dot {\bf q}) dt =
  S(t)=(I_1 \omega_1+I_2 \omega_2)t,
\eeq                                     
so one immediately obtains the semiclassical energy as
\beq
  E=\frac{\partial S}{\partial t}=I_1 \omega_1+I_2 \omega_2.
  \label{energy}
\eeq
In our case we took the quantized values for $I_1$ and $I_2$ and taken into
account that for billiard systems the angular frequencies are of the form
\beq
\omega_m=\lambda_m \sqrt{E},
\eeq
where $\lambda_m$ are frequencies dependent again only upon the geometry of
the chosen torus.

In principle one could obtain the semiclassical energy without considering
the angular frequencies $\omega_i$ by simply finding the appropriate torus 
through the quantization conditions \ref{quantization} and reading its energy.
In a KAM system such as ours, however, these quantization conditions may be 
only approximately fullfiled due to the fine structure of the phase space. In
contrast to the actions $I_i$ the frequencies $\omega_i$ do not depend as
strongly on the initial conditions (close to the main periodic orbit of an 
island of stability the transversal frequency typically does not vanish but is
characteristic of the periodic orbit). So by taking the quantized values for 
actions $I_i$ and the numerical values for $\omega_i$ in equation 
(\ref{energy})
the best estimate of the semiclassical energy is obtained.

Once the proper torus has been found, we proceed to the second step. 
We need to span the semiclassical
wavefunction (\ref{semiclassical}) on this torus. 
Again we deal only with a trajectory, so we must find means
of representing the wavefunction with it. We may write the wavefunction in the
form
\beq
  \psi_j({\bf q})=\lim_{T\to\infty}1/T\int_0^Tdt \delta({\bf q}_{cl}(t)-{\bf
q})D_j({\bf q}_{cl}(t))\exp\left(\frac{i}{\hbar}S_j({\bf q}_{cl}(t)+\phi_j
\right),
  \label{construction}
\eeq
where ${\bf q}_{cl}(t)$ is the classical trajectory in configuration space.
This definition is appropriate only in the sense of integrals of the
wavefunction over configuration space. 
We can hence determine the function $D_j$ by integrating the
wavefunction over a small volume $V^{\prime}$,
\beq
  \int_{V^{\prime}} dV \psi_j=\lim_{T\to\infty}D_j({\bf
q})\exp\left(\frac{i}{\hbar}S_j({\bf q})+\phi_j\right)\frac{T^{\prime}}{T},
 \label{integral}
\eeq
where $T^{\prime}$ is the time the trajectory spends in the volume
$V^{\prime}$. This time is proportional to the density of trajectories $A_j^2$
in the volume $V^{\prime}$,
\beq
 T^{\prime}=T A_j^2({\bf q}) V^{\prime}.
\eeq
If we want the expression (\ref{integral}) to be consistent with
(\ref{semiclassical}),
\beq
  D_j=\frac{1}{A_j} \label{density}
\eeq
must hold.

We still have to account for the phase shifts when the trajectory traverses
from one projection of the torus onto another. This can be done hand in hand
with the estimation of the density of trajectories $A_j^2$. Again we
start a trajectory ${\bf q}_{cl}^{\prime}$ close to the original one (more
trajectories are needed in more than two degrees of freedom). 
We could again use the
monodromy matrix approach, but the same criticism applies here as in the case
of finding the appropriate tori. Let us imagine a bundle of
trajectories inside a small
parallelogram spanned by the three points ${\bf q}_{cl}$, ${\bf
q}_{cl}^{\prime}$ and ${\bf q}_{cl}+\dot {\bf q}_{cl}\delta t$, where all 
three points lie on the given torus. The area of the parallelogram is
given by the absolute value of
\beq
  {\bf P}=\dot {\bf q}_{cl}\times ({\bf q}_{cl}^{\prime}-{\bf q}_{cl})
\delta t.
\eeq
As we are interested only in the relative sizes of this parallelogram, we 
may set $\delta t=1$.
The reciprocal value of this area is proportional to 
the density of trajectories,
\beq
 A_j^2=\alpha/|P|.
\eeq
The value of $\alpha$ does not change along the trajectory and can be
subsequently determined by the normalization of the semiclassical wavefunction.
Whenever the value of $P$ changes its sign, the trajectory has encountered a
caustic and has passed from one projection $j$ of the torus to another one $k$.
At this point the wavefunction acquires a phase shift 
\beq
  \phi_k=\phi_j-\pi/2.
\eeq
The phase shift is equal to $\pi$ if a hard wall is encountered.

The construction (\ref{construction}) is, in its original form, of course
unsuitable for numerical computation. We must substitute the $\delta$ function
in the integral by a function of finite width. The other important reason for
doing so is to again smooth the classical behaviour which can have a far more
detailed structure than the quantum mechanics can yet resolve. This width may 
not necessarily need be isotropic. 
A good estimate for it is again to be of the order 
of $1/n_i$ in coordinates $\theta_i$ projected onto configuration space.

We used a Gaussian for the wide delta function. If we adjusted the width and
the amplitude of the
Gaussian so that it followed the classical density of trajectories, 
a remarkable
similarity of our method with that of Heller's wavepacket approach was
observed (Heller 1991). There are, however, two important differences. The 
wavepacket approach starts with a wavepacket with the expected value of 
energy equal to that of the exact
wavefunction and then tries to construct its semiclassical approximation. Our
approach is independent of the exact eigenstate. All the information we have to
supply is the quantum numbers and the geometrical properties of the 
(projection of the) torus,
obtaining from them both the semiclassical energy (\ref{energy}) and the
wavefunction. The other important difference is that the wavepacket approach
relies on the monodromy matrix of the trajectory, the use of which can be
questionable to obtain semiclassical wavefunctions in mixed type systems 
due to the fine structure of the classical phase space, as was already pointed
out before. 

We show  three examples of the regular states in the figures \ref{semi1} to 
\ref{semi3}. For 
each of the states we present the exact 
numerical quantum probability density(top left), the probability density 
of its semiclassical approximation (top right), 
the classical density of trajectories on the appropriate torus (bottom left)
and the smoothed projection of the exact Wigner function (bottom right).  
The semiclassical wavefunctions shown are remarkable as they possess
all of the features of their exact counterparts that are larger than the 
appropriate wavelength. Note that for each torus 
there are two characteristic wavelengths since there are two 
quantum numbers associated with it. As it happens in our case, the two 
wavelengths can be of different orders of magnitude.

There were, however, some 'regular' states that we were 
unable to semiclassically reproduce. The first class of these states can be
described as localized chaotic states since their Wigner function clearly shows
that they lie in the chaotic region, yet very close to an island of stability,
 which gives them a regular appearance. The other class
are the states whose Wigner transforms lie in the regular regions, but where
the primary tori have already been destroyed by the perturbation 
and now form secondary tori
interwoven by small regions of chaotic motion, and even the smoothing of
classical dynamics, as described above, fails.

The accuracy of the semiclassical energies that we were indeed able to
reproduce may seem remarkable, since the error is approximately $5$ units of
energy at the energies around $2\cdot 10^7$. Such accuracy, however, is still
insufficient to perform short range spectral statistics since the mean level
spacing in our system is approximately $8$ units of energy. This experience
is of course in agreement with the proposition and conclusion that the
semiclassical methods (to the leading order) cannot resolve the energy
spectra within the vanishing fraction of the mean level spacing, and
also not the structures of the wavefunctions smaller than de Broglie
wavelength (Prosen and Robnik 1993a, Robnik and Salasnich 1997).

\subsection{Irregular states}

While for the regular states it was quite straightforward to find their
semiclassical approximations, the nature of irregular states is very much
different. The chaotic component of a system does not possess any obvious
structure. While Gutzwiller's approach can yield the properties of a quantum
system by a summation over all periodic orbits of its classical counterpart,
the relevance of examining individual chaotic states becomes questionable.
These states are very sensitive to small perturbations of the system, so in any
physical system the individual features of the states are lost 
when the effective Planck's constant tends to $0$. The features that are
insensitive
to small perturbations are, however, the statistical properties of spectra and
eigenstates.

One measure of statistical properties of the wavefunctions is the wavefunction
autocorrelation function,
\beq
  C({\bf q}, {\bf x})=\frac{\langle \psi^{\dagger}({\bf q}^{\prime}-{\bf x}/2)
  \psi({\bf q}^{\prime}+{\bf x}/2)
  \rangle_{{\bf q}^{\prime}\in \epsilon({\bf q})}}{\langle\psi^{\dagger}
  ({\bf q^{\prime}})
  \psi({\bf q}^{\prime})\rangle_{{\bf q}^{\prime}\in \epsilon({\bf q})}}.
  \label{autocorrelation}
\eeq
The area of averaging $\epsilon({\bf q})$ close to the point ${\bf q}$ 
should be taken such that its linear size is many wavelengths across, however
small enough that the local properties of classical mechanics within it are
largely uniform.

If one takes the Fourier transform of the Wigner function (\ref{Wigner}),
it is easy to show that
\beq
  \int W({\bf q},{\bf p}) \exp(i {\bf p}\cdot {\bf x}/\hbar) d^{N} {\bf p}=
  \psi^{\dagger}({\bf q}-{\bf x}/2)\psi({\bf q}+{\bf x}/2). 
  \label{fourier} 
\eeq                     
By knowing the Wigner function of an eigenstate,
it is then possible to use this result to calculate
its autocorrelation function.
                                     
According
to the principle of uniform semiclassical condensation, the Wigner
function of any chaotic state should uniformly condense on the whole chaotic
component when the effective $\hbar$ tends to $0$. 
Let us limit ourselves only to the cases of the Hamiltonians with an
isotropic dependence upon ${\bf p}$.
We can write the semiclassical Wigner function in the
form of a conditional delta function 
\beq
  W_{{\cal D}_i}
  ({\bf q},{\bf p})=\alpha \delta(\{{\bf q},{\bf p}\} \in {\cal D}_i;
 E-H({\bf q},p))
\eeq                                
where ${\cal D}_i$ 
denotes a chaotic component and $\alpha$ is the normalization
constant. The Fourier transform (\ref{fourier}) of this Wigner function is
\beq
 \int 
W_{{\cal D}_i}({\bf q},{\bf p}) p^{N-1} \exp(i {\bf p}\cdot {\bf x}/\hbar) 
 dp d\Omega_p =
 \alpha \frac{p({\bf q})^{N-1}}{\frac{\partial H}{\partial p}({\bf q},p({\bf 
 q}))}\int_{\Omega_p \in {\cal D}_i({\bf q})} 
 d\Omega_p \exp(i {\bf p}\cdot {\bf x}/\hbar),
\eeq
where $p({\bf q})$ denotes the absolute value of momentum at the point ${\bf 
q}$. The integration over the spatial angle $\Omega_p$ 
is performed along all the directions of 
momentum that constitute the chaotic component ${\cal D}_i$ 
at the point ${\bf q}$.
The autocorrelation function is then equal to
\beq
  C_{{\cal D}_i}({\bf q}, {\bf x})=\frac{\langle \int_{\Omega_p \in 
{\cal D}_i({\bf q}^{\prime})} 
 d\Omega_p \exp(i {\bf p}\cdot {\bf x}/\hbar)\rangle_{{\bf q}^{\prime}
 \in \epsilon({\bf q})}}{\langle\int_{\Omega_p \in 
{\cal D}_i({\bf q}^{\prime})} 
 d\Omega_p \rangle_{{\bf q}^{\prime}\in \epsilon({\bf q})}}.
\eeq
The averaging area should again stretch across many wavelengths.

If the chaotic component is equal to the whole energy surface, as is the case 
in completely ergodic systems, in the case of two degrees of freedom 
one obtains
the well known Berry's result (Berry 1977)
\beq
  C_{\mathrm ergodic}({\bf x})=J_0(p({\bf q}) r/\hbar), \  r=|{\bf x}|.
  \label{Berry}
\eeq  
However, when the system is of the mixed type the autocorrelation function 
ceases to be isotropic and acquires contributions of higher order Bessel 
functions. We can obtain these contributions by rewriting 
integrals $\int_{\phi_p \in {\cal D}_i} 
f(\phi_p) d\phi_p$ by integrals of the characteristic function $\int 
\chi_{{\cal D}_i}(\phi_p) f(\phi_p) d\phi_p$, 
where in two degrees of freedom the spatial 
angle is replaced by a simple angle $\phi_p$ and $f(\phi_p)$ is an arbitrary 
function of $\phi_p$. If we write the characteristic 
function as a Fourier series,
\beq
  \chi_{{\cal D}_i}({\bf q};\phi_p)=\sum_{m=-\infty}^{\infty}
  \kappa_m^{{\cal D}_i}({\bf q})
   \exp(i m \phi_p),
\eeq 
it is quite straightforward to show by using the integral representations of 
the Bessel functions that
\beq
 C_{{\cal D}_i}({\bf q}, {\bf x})=\frac{\langle \sum_{m=-\infty}^{\infty}
 \kappa_m^{{\cal D}_i}({\bf q^{\prime}}) 
 i^m J_m(p({\bf q^\prime}) r/\hbar)\exp(i m 
 \phi_x)
 \rangle_{{\bf q}^{\prime}
 \in \epsilon({\bf q})}}{\langle
 \kappa_0^{{\cal D}_i}({\bf q^{\prime}}) 
 \rangle_{{\bf q}^{\prime}\in \epsilon({\bf q})}},
\eeq
where $\phi_x$ is the polar angle of the vector ${\bf x}$.

As in the case of the regular states, numerically we cannot deal with 
components of phase space but with trajectories. So we start a trajectory 
within the chaotic component ${\cal D}_i$ 
and not the direction of its momentum 
$\phi_p^j({\bf q})$ 
at each passage $j$ through the neighbourhood of the point ${\bf q}$. 
The averaged characteristic function for this neighbourhood 
can then be represented as
\beq
  \chi_{{\cal D}_i}({\bf q};\phi_p)/
  \kappa_0^{{\cal D}_i}({\bf q})=\frac{ \sum_{j=1}^n 
  d_j \delta(\phi_p-\phi_p^j)}{\sum_{j=1}^n d_j}, \label{characteristic}
\eeq 
where $d_j$ are the lengths covered by the particle in the averaging 
neighbourhood at the passage $j$, and $n$ is the number of passages. 
One can check 
that this method tends to the proper angular distribution $\chi_{{\cal D}_i}$ 
as
$n\to\infty$ for any shape of the averaging neighbourhood, if one assumes 
the homogeneity of trajectories at a given angle (which is true for a small 
enough neighbourhood, over which the classical phase space picture does not 
vary). For any incident angle $\phi$ the conditional expected value of a
contribution to the equation (\ref{characteristic}) is proportional to 
$\int d_j dx$, where $x$ is the homogeneously distributed impact parameter (the
direction perpendicular to the incident angle).
This integral gives just the area of the averaging neighbourhood and is clearly
the same for all incident angles. 

If the area of averaging is large enough so
that the variations in the classical phase space picture become important, the
above procedure is still valid as long as the value of the 
momentum does not change
appreciably (as is the case in our billiard system, where between the
bounces the 
momentum remains constant). One can then imagine the large averaging area as
being cut
into smaller ones within which the above assumptions still hold true.

We compared the autocorrelation functions for a few chaotic states with the 
semiclassical prediction in figures \ref{chaos1} to \ref{chaos3}. 
The averaging area $\epsilon({\bf q})$ 
was taken as a circle of radius $0.2$ around the point $(x,y)=(0.65,0)$
(the coordinates are as defined in equation (1)).
It was taken the same for both the semiclassical 
prediction and for the numerical results. The averaging 
radius was taken quite large in 
order to reduce the localization properties of the wavefunctions, which are 
still apparent at the values of effective $\hbar$ that we were able to obtain. 
But this radius still has to be taken small enough in order not to completely
smooth out the classical dynamics. The agreement with the semiclassical 
prediction is quite good particularly in figure 14. In all cases it clearly deviates from 
the Berry's prediction for fully ergodic systems (\ref{Berry}), 
as it must for mixed systems,  and tends 
towards our semiclassical result. Although in some cases there are amplitude 
deviations from our prediction, in all of the plots the phase of the
numerical correlation function matches the phase of its semiclassical
prediction and is significantly different from the phase predicted for the 
fully ergodic case.

\section{Discussion and conclusion}

\noindent As already presented by Prosen and Robnik (1993c), the classification
of states into regular and irregular ones is well founded when the effective
Planck's constant tends to $0$. Its theoretical foundation is the principle of
uniform semiclassical condensation of Wigner functions of eigenstates (Robnik 
1988, 1998). This separation is not strictly a semiclassical
phenomenon since even some of the lowest levels in our catalogue
of low lying states can be classified as either regular or irregular.
In the high energy catalogue of states each state can easily be classified as
either chaotic or regular, with only one notable exception, where two close
lying states are a superposition of a regular and an irregular state. These
exceptions are expected to disappear with higher energies. While the states can
be separated with respect to classical dynamics, the chaotic states in our high
energy catalogue still exhibit the
phenomenon of dynamical quantal 
localization. Their Wigner functions are not uniformly extended
over the whole chaotic component, but are significant only on a part of it.
This localization is expected to disappear at sufficiently small effective
$\hbar$, when the quantum mechanical break time
$t_{break}=\hbar/\Delta E$, where $\Delta E$ is the mean level spacing, becomes
longer than the time for a typical trajectory to explore the 
whole chaotic component (diffusion time).

The most important part of this work is of course the semiclassical analysis of
states. We were able to reconstruct both the semiclassical wavefunction and the
semiclassical energy of the regular states by using the EBK quantization. Since
our system is of a KAM type, most of the resonant tori are destroyed, forming 
smaller islands of stability interwoven with chaotic components. The classical
mechanics thus shows a rich structure that quantum mechanics at a fixed value
of $\hbar$ is still unable to resolve. In order to obtain the regular
semiclassical wavefunctions, we had to appropriately smooth out this fine
classical behaviour as explained in the section \ref{analysis}. 

We were unable to predict
the individual properties of chaotic states. We made a step forward, however,
in describing their semiclassical statistical properties. We obtained a
semiclassical prediction for the autocorrelation function of their
wavefunction, which differs from the one for fully ergodic systems as it 
is not isotropic. The numerical results confirm this prediction,
although there are still localization phenomena at the currently attainable
effective $\hbar$ that cause deviations from it.

One aspect that needs to be investigated further 
is the localization properties of the chaotic states. As Casati and Prosen
(1998) show in the example of a fully chaotic stadium billiard, in 
the diffusive regime ($\epsilon$-stadium, having very large ergodic time),
the quantum
diffusion is stopped by the cantori in phase space leading to localization. 
How these and similar ideas translate to the case of a mixed type system 
remains so far an open question.

\section*{Acknowledgments}
We thank Dr. Toma\v z Prosen for assistance and advise with some computer
programs. This work was supported by the Ministry of Science and Technology
of the Republic of Slovenia and by the Rector's Fund of the
University of Maribor.

\newpage

\section*{References}

\setlength{\parindent}{0cm}
Berry M V 1977, {\it J. Phys. A: Math. Gen.} {\bf 10} 2083
\\\\
Berry M V 1983 {\it Chaotic Behaviour of Deterministic Systems (Proc. NATO ASI 
Les Houches Summer School} ed G Iooss, R H G Helleman, R Stora (Amsterdam: 
Elsevier) p 171
\\\\
Berry M V and Robnik M 1984 {\it J. Phys. A: Math. Gen.} {\bf
17} 2413
\\\\
Berry M V and Wilkinson M 1984 {\it Proc. Roy. Soc. London A} {\bf 392} 15
\\\\
Bogomolny E 1988 {\it Physica} {\bf 31D} 169
\\\\
Casati G and Prosen T 1999 {\it Phys. Rev. E} {\bf 59} 2516
\\\\
Gutzwiller M 1990, {\it Chaos in Classical and Quantum
Mechanics}, Springer New York
\\\\
Heller E J 1984 {\it Phys. Rev. Lett.} {\bf 53} 1515
\\\\
Heller E J 1986 {\it Lecture Notes in Physics} {\bf 263} 162
\\\\
Heller E J 1991 {\it Chaos and Quantum Physics (Proc. NATO ASI Les Houches
Summer School} ed M-J Giannoni, A Voros and J Zinn-Justin (Amsterdam: Elsevier)
p 547
\\\\ 
Lazutkin V F 1981 {\it The Convex Billiard and the Eigenfunctions of the 
Laplace Operator} (Leningrad: University Press) (in Russian)
\\\\
Lazutkin V F 1991 {\it KAM Theory and Semiclassical Approximations to 
Eigenfunctions} (Heidelberg: Springer)
\\\\
Li Baowen and Robnik M 1994 {\it J. Phys. A: Math. Gen.} {\bf
27} 5509
\\\\
Li Baowen and Robnik M 1995a {\it J. Phys. A: Math. Gen.} {\bf 28}
2799
\\\\
Li Baowen and Robnik M 1995b {\it J. Phys. A: Math. Gen.} {\bf 28}
4483
\\\\
Percival I C 1973 {\it J. Phys. B: At. Mol. Phys.} {\bf 6} L229
\\\\
Prosen T 1995 {\it J. Phys. A: Math. Gen.} {\bf 28} L349
\\\\
Prosen T 1996 {\it Physica D} {\bf 91} 244
\\\\
Prosen T 1998 {\it J. Phys. A: Math. Gen.} {\bf 34} 7023
\\\\
Prosen T and Robnik M 1993a {\it J. Phys. A: Math. Gen.} {\bf 26} L37
\\\\
Prosen T and Robnik M 1993b {\it J. Phys. A: Math. Gen.} {\bf 26} 2371
\\\\
Prosen T and Robnik M 1993c {\it J. Phys. A: Math. Gen.} {\bf
26} 5365
\\\\
Prosen T and Robnik M 1994 {\it J. Phys. A: Math. Gen.}
{\bf 27} 8059
\\\\
Prosen T and Robnik M 1999 {\it J. Phys. A: Math. Gen.} {\bf 32} 1863
\\\\
Robnik M 1983 {\it J. Phys. A: Math. Gen.} {\bf 16} 3971
\\\\
Robnik M 1984 {\it J. Phys. A: Math. Gen.} {\bf 17} 1049
\\\\
Robnik M 1988 in {\em Atomic Spectra and Collisions in External Fields}
Eds. K.T. Taylor, M.H. Nayfeh and C.W. Clark (New York: Plenum Press)
pp251-274
\\\\
Robnik M 1989 {\em Bound-State Eigenfunctions of Classically
Ergodic Hamilton Systems: A Theory of Scars} Preprint Institute
for Theoretical Physics, University of California Santa Barbara,
unpublished
\\\\
Robnik M 1998 {\it Nonlinear Phenomena in Complex Systems} {\bf
1} 1
\\\\
Robnik M and Salasnich L 1997 {\it J. Phys. A: Math. Gen.} {\bf 30} 1711
\\\\
Robnik M and Veble G 1998 {\it J. Phys. A: Math. Gen.} {\bf 31} 4669
\\\\
Vergini E and Saraceno M 1995 {\it Phys. Rev. E} {\bf 52} 2204
\\\\
Voros A 1979 {\it Lecture Notes in Physics} {\bf 93} 326

\newpage

\begin{figure}
\caption{The SOS section of the 
 main chaotic component of the $\lambda=0.15$ billiard. The
coordinate $x$ here is shifted by $\lambda$ to the right so that
$x\in[-1,+1]$, and $x=0$ corresponds to $x=\lambda$ of equation (1).}
\label{sos}
\end{figure}

\begin{figure}
\caption{The shortest periodic orbits of the $\lambda=0.15$ billiard (explained
in text). The stable orbits are shown with full and the unstable ones with 
dashed lines.}
\label{orbits}
\end{figure}

\begin{figure}
\caption{A selection of eigenstates corresponding to the stable 
periodic orbits 
1, 2, 3, 4 
in the figure \ref{orbits} from top to bottom, respectively. In the top 
half of each plot we show the eight equally spaced contours 
of the probability density 
from $0$ to its maximum value. In the bottom half we show the nodal 
lines of each wavefunction.}
\label{lowstatesstable}
\end{figure} 

\begin{figure}
\caption{A selection of eigenstates corresponding to the stable 
periodic orbits 
5 and 6 
in the figure \ref{orbits} in the top and bottom row, respectively. 
In the top 
half of each plot we show the eight equally spaced 
contours of the probability density 
from $0$ to its maximum value. In the bottom half we show the nodal 
lines of each wavefunction.}
\label{lowstatesunstable}
\end{figure}

\begin{figure}
\caption{The corresponding smoothed projections of the 
Wigner functions for the wavefunctions in the figure
\ref{lowstatesstable}. The contours are spaced in ten intervals from $0$ to
the maximal value. The negative 
value contours are plotted at the same spacing but with thinner lines. We do 
not plot the zero level contour due to many oscillations of the Wigner 
functions when it is close to zero. As in all subsequent Wigner plots, the 
momentum $p_x$ is measured in units of $p^\prime$, which is related to 
the energy of the  eigenstate by $E=p^{\prime2}$. The locations of the 
relevant classical periodic orbits are marked by the "bullets". 
The coordinate $x$ here is shifted by $\lambda$ as in figure 1.}
\label{lowstateswignerstable}
\end{figure}

\begin{figure}
\caption{The corresponding smoothed projections of the 
Wigner functions for the wavefunctions in the figure
\ref{lowstatesunstable}. The plotting method is the same as in figure 
\ref{lowstateswignerstable}.}
\label{lowstateswignerunstable}
\end{figure}

\begin{figure}
\caption{An example of the probability density for 
 a regular (top left) and an irregular (top right) state with
the corresponding smoothed projections of their Wigner functions shown below,
with the same plotting methods as used in corresponding previous plots.
In the phase space plots (bottom row) the coordinate $x$ is
shifted by $\lambda$ as in figure 1. The "bullets" mark the
location of the stable period 3 periodic orbit, which is the
skeleton of the quantized invariant torus.}
\label{twostates}
\end{figure}

\begin{figure}
\caption{The probability density (16 equally spaced contours 
from 0 to the maximum 
value) for a pair of close lying states
($k^2=20421106.7347$ left and  $k^2=20421107.0691$ right)
that are superpositions of a regular and
an irregular state.} 
\label{pair}      
\end{figure}                                

\begin{figure}
\caption{The probability density for a 
 regular state with $k^2=20420831.0603$ (top left), its
 semiclassical approximation with $k_{sc}^2=20420828.18$ (top right), 
 with the quantum numbers on the torus being $n_1=6752$ and $n_2=37$ 
 (16 equally spaced contours from $0$ to the maximum value in both cases).
 In the bottom row we show the classical density 
with 20 contours from 0 to the maximum of the appropriate torus 
 (left) and the
 smoothed projection of the exact Wigner function (right), with ten contours
 from $0$ to the maximum value. In the phase space plot (bottom right)
the coordinate $x$ is shifted by $\lambda$ as in figure 1.}\label{semi1}
\end{figure}

\begin{figure}
\caption{The same as in figure \ref{semi1} but for the state
 with $k^2=20421002.7443$ and its semiclassical approximation with 
  $k_{sc}^2=20420999.46$, $n_1=2861$ and $n_2=20$.}\label{semi2}
\end{figure}
\begin{figure}
\caption{The same as in figure \ref{semi1} but for the state
 with $k^2=20421387.1741$ and its semiclassical approximation with 
  $k_{sc}^2=20421385.96$, $n_1=8648$ and $n_2=1$.}\label{semi3}
\end{figure}

\begin{figure}
\caption{In the top row we show the probability density 
 for the chaotic state with $k^2=20420756.1273$
 (left, 8 contours) 
 with the smoothed projection of its Wigner function (right, 10 contours). 
The circle of radius $0.2$ centered at $x=0.5$ is the region of averaging.
In the phase space
plot (top right) the coordinate $x$ is shifted by $\lambda$ as in
figure 1.
In the bottom row we plot the wavefunction autocorrelation function 
(averaged over the small circle as explained in text)
in the $x$ (left) and $y$ (right) directions.}\label{chaos1}
\end{figure}

\begin{figure}
\caption{The same as in figure \ref{chaos1} but for the state with 
$k^2=20421005.3834$.}\label{chaos2}
\end{figure}

\begin{figure}
\caption{The same as in figure \ref{chaos1} but for the state with 
$k^2=20421262.6667$.}\label{chaos3}
\end{figure}


\end{document}